\documentclass[%
aip,
rsi
amsmath,amssymb,
reprint,%
]{revtex4-2}

\usepackage{graphicx}% Include figure files
\usepackage{dcolumn}% Align table columns on decimal point
\usepackage{bm}% bold math
\usepackage[mathlines]{lineno}% Enable numbering of text and display math
\usepackage{textgreek}
\usepackage{amsmath}
\usepackage{ulem}
\newcommand{\Eq}[1]{Eq.\,\ref{#1}}% \Eq{abc}
\newcommand{\Fig}[1]{Fig.\,\ref{#1}}% \Fig{abc}
\newcommand{\Sec}[1]{Sec.\,\ref{#1}}% \Sec{abc}
\newcommand{\Tab}[1]{Table \,\ref{#1}}% \Tab{abc}
\newcommand{\be}{\begin{equation}}
\newcommand{\ee}{\end{equation}}
\newcommand{\bea}{\begin{eqnarray}}
\newcommand{\eea}{\end{eqnarray}}
\newcommand{\bd}{\begin{displaymath}}
\newcommand{\ed}{\end{displaymath}}
\newcommand{\ben}{\begin{enumerate}}
\newcommand{\een}{\end{enumerate}}

\newcommand{\Onlinecite}[1]{Ref.\,\onlinecite{#1}} % \Onlinecite{abc}
\newcommand{\nm}{\mathrm{nm}} %% nanometer

\newcommand{\ca}{c/$a$\ }

\newcommand{\nuc}{\ensuremath{\nu_\mathrm{c}}} %% reduced real resonance frequency
\newcommand{\nut}{\ensuremath{\nu_\mathrm{t}}} %% target frequency

\newcommand{\omcc}{\ensuremath{\tilde{\omega}_\mathrm{c}}} %% complex resonance angular frequency
\newcommand{\omc}{\ensuremath{\omega_\mathrm{c}}} %% real resonance angular frequency
\newcommand{\gamc}{\ensuremath{\gamma_\mathrm{c}}} %% Imaginary part of complex angular resonance frequency
\newcommand{\gamcr}{\ensuremath{\Gamma_\mathrm{c}}} %% reduced linewidth FWHM

\newcommand{\tauc}{\ensuremath{\tau_\mathrm{c}}} %% photon lifetime of resonance

\newcommand{\lamc}{\ensuremath{\lambda_\mathrm{c}}} %%resonant wavelength
\newcommand{\omt}{\ensuremath{\omega_\mathrm{t}}} %% target frequency

\newcommand{\br}{\ensuremath{\mathbf{r}}} %% position
\newcommand{\Vm}{\ensuremath{V_\mathrm{m}}} %% mode volume

\newcommand{\bk}{\ensuremath{\mathbf{k}}} %% wavevector
\newcommand{\bkip}{\ensuremath{\mathbf{k}_{||}}} %% in-plane wavevector

\newcommand{\nme}{\ensuremath{n_\mathrm{m}}} %% medium refractive index
\newcommand{\nsl}{\ensuremath{n_\mathrm{s}}} %% slab refractive index

\newcommand{\dme}{\ensuremath{d_\mathrm{m}}} %% medium thickness
\newcommand{\dpml}{\ensuremath{d_\mathrm{p}}} %% PML thickness
\newcommand{\dsl}{\ensuremath{d_\mathrm{s}}} %% slab thickness
\newcommand{\rh}{\ensuremath{R^\mathrm{h}}} %% hole radius

\newcommand{\xyh}{\ensuremath{\mathbf{r}^\mathrm{h}}} %% hole position

\newcommand{\parm}{\ensuremath{\mathbf{p}}} %% optimisation parameters

\newcommand{\parmd}{\ensuremath{\mathbf{\Delta}}} %% change of optimisation parameters

\newcommand{\xyhde}{\ensuremath{\Delta_\mathbf{r}}} %% hole position change

\newcommand{\rhde}{\ensuremath{\Delta_R}} %% hole radius change

\newcommand{\Fgr}{\ensuremath{\mathbf{G}}} %% gradient of cost function

\newcommand{\Fgrh}{\ensuremath{\hat{\mathbf{G}}}} %% normalised gradient of cost function

\usepackage{xr}
\begin{document}
	
%\preprint{AIP/123-QED}

%\title[Sample title]{Sample Title:\\with Forced Linebreak\footnote{Error!}}% Force line breaks with \\
%\thanks{Footnote to title of article.}
\title{Designing low-loss cavities across the band-gap of photonic crystal slabs}

\author{Nadhia Monim}
\affiliation{ 
	School of Biosciences, Cardiff University, Sir Martin Evans Building, Museum Avenue, Cardiff, CF10 3AX%\\This line break forced with \textbackslash\textbackslash
}%

\author{Wolfgang Langbein}%
\email{langbeinww@cardiff.ac.uk}
\affiliation{ 
	School of Physics and Astronomy, Cardiff University, The Parade, Cardiff CF24 3AA, United Kingdom%\\This line break forced with \textbackslash\textbackslash
}%

\author{Francesco Masia}
\email{masiaf@cardiff.ac.uk}
\affiliation{ 
	School of Biosciences, Cardiff University, Sir Martin Evans Building, Museum Avenue, Cardiff, CF10 3AX%\\This line break forced with \textbackslash\textbackslash
}%

\date{\today}% It is always \today, today,
	
\begin{abstract}
Photonic crystal cavities (PCCs) are defects in host photonic crystals (PCs) which create bound states in the PC band gap. These bound states are resonant states of the electromagnetic field with a complex resonance frequency and can have very small mode volumes. PCCs are attractive for a variety of applications, from cavity quantum electrodynamics to biosensing. A PC slab geometry is advantageous given its superior manufacturability compared to three-dimensional crystals, and the accessibility of the surface allows sensing and coupling. However, the emission into the half spaces above and below the slab limits the bound state lifetime.  Controlling this emission is thus crucial for applications, most of which benefiting from a long lifetime. A range of methods to find defect geometries suppressing the emission to increase the lifetime have been demonstrated in the past. However, they do not cater for a designed resonant frequency covering a wide addressable range, as needed for multiplexed devices. Here, we demonstrate a design method controlling both resonance frequency and emission, by minimising a cost function including both losses and target frequency. We show applications on PCCs in GaAs PC slabs immersed in water, relevant for biosensing. The reduced refractive index contrast in these structures compared to previously studied PCCs embedded in vacuum renders the emission suppression more challenging. We optimize the quality factor of a standard L3 cavity from 1000 to $10^4-10^5$, with an addressable resonance frequency range covering 12\% relative bandwidth, spanning more than half of the band gap. We furthermore report optimised structures of H1 cavities, and provide the optimisation code for widespread use.
\end{abstract}

%\keywords{pcc}%Use showkeys class option if keyword
\maketitle

\section{Introduction}
Photonic crystals (PCs) are photonic structures having spatially periodic optical material properties. They can develop a photonic band gap, i.e. a range of frequencies over which light cannot propagate in the structure.\cite{JoannopoulosBook08} By tuning the properties of the periodic pattern, the size and position of the photonic band gap can be controlled. Two-dimensional (2D) PCs can be obtained by etching a periodic pattern of holes in a membrane of semiconducting material.\cite{KraussN96} 
By introducing a defect in the PC, breaking the translational symmetry of the periodic pattern, bound states of frequencies within the band gap can be formed, which are the modes of the photonic crystal cavity (PCC) formed by the defect.\cite{LalanneLPR08} These devices have a range of potential application in photonics and quantum communication such as filters,\cite{BazianPNC20} optical modulators,\cite{PanuskiNPho22} and low threshold lasers.\cite{EllisNPho11}
 
The bound states are resonant states (RS) with a complex angular frequency $\omcc=\omc+i\gamc$, composed of their angular frequency \omc\ and a field decay rate \gamc\ quantifying losses. The photon lifetime $\tauc$ in the resonant state is given by $\tauc=1/(2\gamc)$, determining the RS quality factor $Q=\omc\tauc$. The large lifetime of a high $Q$ RSs allows for higher sensitivity in sensing and strong coupling in cavity quantum electrodynamics.

The losses of RSs in 2D PCCs are given by material absorption, radiative losses into 3D space, and in-plane losses due to the finite extension of the PC surrounding the defect.  
The material losses depend mostly on the frequency and the materials used, and the photonic design can help by minimising the electromagnetic energy inside the absorbing material, as in slot waveguides.\cite{LiPTL16} In-plane losses can be controlled by simply extending the PC size, and choosing the frequency well within the photonic bandgap in order to decrease the in-plane mode field extension.

The radiative losses due to emission into the half-spaces to either side of the 2D PC are depending on the defect structure, and are determined by the RS field in wavevector (\bk) space. Any field amplitude at a wavevector \bk\ with a component \bkip\ in the PC plane within the radiative cone of the surrounding medium, $|\bkip|<\omc\nme/c$, with the speed of light $c$, creates such emission. Due to fabrication constraints, 2D PCs are often made of a periodic arrangement of holes in a slab, where an hexagonal lattice is preferred compared to a square lattice as it provides a larger bandgap. The simplest way to create defects in these 2D PCs is to omit, shift, or resize holes. Simple defects are created by omitting holes, and are called for example H1 for a single hole or L3 for three adjacent holes in a line.

Previous literature has reported a reduction of radiative losses by altering the position and size of the holes surrounding the PCC.\cite{AkahaneN03,MinkovSR14,VascoSR21,LiNanoM22,GranchiACSP23}
The corresponding increase of $Q$ has been experimentally validated.\cite{AkahaneN03,LaiAPL14} Notably, the reported methods do not explicitly control the frequency \omc\ to match a target frequency \omt. However, for applications of PCCs also \omc\ needs to be designed, for example when using them as transducing elements in biosensors.\cite{ZhangSAAP15} Once functionalised with receptor molecules, the attachment of analytes from the environment to the device surface can be detected via the local change of polarisability probed by the evanescent field of the RS. As a result, the RS frequency shifts from its design value by an amount proportional to the number of attached analyte molecules, so that the detection limit is proportional to the linewidth, and thus $1/Q$. Multiplexed detection can be achieved by functionalising PCC transducers with different receptors, and they can be independently read if their RSs have sufficiently different frequencies, similar to the microwave readout of kinetic inductance detector arrays.\cite{McHughRSI12} Therefore, not only has $Q$ to be maximised, but also \omc\ needs to be controlled to a given target.

In this work, we demonstrate an algorithm which optimises the position and size of the holes surrounding the omitted holes to achieve a mode at a controlled target frequency with the maximised $Q$. The algorithm is based on a gradient descent approach, which minimises a cost function comprising both the detuning from a target frequency, and the decay rate. 

As application example we consider PCCs made from a GaAs slab immersed in water at wavelengths $\lamc=2\pi c/\omc$ around 1.3\,\textmu m. Owing to its direct band gap, the (In,Ga,Al)As system has the potential to integrate excitation (PCC lasers), sensing (PCCs), and detection (photoelectric diodes) elements, motivating this material choice.  We show an enhancement of $Q$ by about two orders of magnitude to $Q>7\times 10^4$ for a L3 cavity over a \lamc\ range of 25\,nm, and by about three orders of magnitude to $Q>5\times 10^4$ for a H1 cavity, over a \lamc\ range of 50\,nm.
We note that due to the lower index contrast of the PCC in water medium, achieving low radiative losses in our structure is more challenging than for PCC in vacuum, for which quality factors above $10^6$ were achieved.\cite{LaiAPL14}

\section{Methods}

To calculate the PC and PCC electromagnetic modes, we use a commercial finite element software (COMSOL) using an eigenmode solver. We first optimise the PC parameters, by determining the hole radius \rh\ and slab thickness \dsl\ maximising the width of the photonic band gap of the GaAs slab in water. The simulation domain is shown in the supporting information (SI) \Fig{S-fig:PCGeom}, defining the lattice constant $a$. The largest band gap for TE modes occurs at $\rh=0.4a$ and $\dsl=0.8a$. However, to obtain a more mechanically robust structure of the membrane suited for fabrication, we use a slightly smaller value of $\rh=0.35a$, which is reducing the bandgap by about 20\%, spanning from $0.255$ to $0.337\,c/a$.  Using a lattice constant of $a=325$\,nm, the nominal radius of the holes is $\rh=0.35a=114$\,nm, and the slab thickness $\dsl = 0.8a = 260$\,nm, yielding a band gap centred around $1.2$\,\textmu m. To compare the optimisation with literature, we have also modelled a PCC made from a Si slab in air with $\dsl=240$\,nm thickness, in a PC with lattice constant $a=437$\,nm and hole radius of $0.35a=153$\,nm, to achieve a resonant mode around $1.55\,$\textmu m. Details of the PC optimisation are discussed in \Sec{S-OptPC}.

For the PCC calculations, we use a diamond shape domain with periodic boundary conditions in plane, as detailed in \Sec{S-sec:domain}. The size of the domain is chosen large enough to neglect coupling between the PCCs in the periodic array. We found that the cavity mode linewidth is stable for lateral sizes larger than $10\,a$ along the short axis, as shown in \Sec{S-sec:domain_size}, and we used $15\,a$ for the optimisations shown here to ensure negligible coupling. The semiconductor slab is covered by a medium layer of $\dme=614$\,nm thickness, followed by a perfectly matched layer (PML) of $\dpml=1228$\,nm thickness, and perfect conductor boundary conditions. The medium thickness was chosen such that the near field has decayed sufficiently to neglect its absorption by the PML as shown in \Sec{S-sec:domain_dm}. Similarly, the PML thickness was chosen such that the reflection of the radiated field is sufficiently suppressed, as shown in \Sec{S-sec:domain_pml}. 

For clarity, we define a unitless frequency $\nuc=\omc a/(2\pi c)$ and full width at half maximum linewidth $\gamcr=\gamc a/(\pi c)$. In the optimisation, we define a cost function $F=\tilde{F}P$, where $P$ is a penalty function to constrain parameters to physically achievable geometries, and 
\be
\tilde{F}=\sqrt{(\nuc-\nut)^2+(\alpha\gamcr)^2},
\ee
with the target frequency \nut\ and the relative weight  $\alpha$ of the linewidth compared to the frequency mismatch. The optimisation parameters \parm\ are the positions $\xyh_i$ and radii $\rh_i$ of selected holes surrounding the cavity, which are changed to minimise $F(\parm)$. For both the L3 and H1 cavity, we choose to retain their mirror symmetries as shown in \Sec{S-PCCParam}. This choice is motivated by the cylindrical symmetry of the radiative cone around the out-of plane axis, suggesting that a higher symmetry of the cavity geometry about this axis is helping the RS field to avoid the radiative cone. Furthermore, these symmetries are reducing the number of optimisation parameters, and thus the computational complexity. The holes optimised in the L3 PCC were selected for their large overlap with the considered M1 cavity mode. Considering computational time and overlap, we chose 20 holes being optimised, described by a total of 16 parameters.  For the H1 PCC, three rings of holes were optimised, comprising 36 holes described by a total of 6 parameters. 

To enforce geometrical and manufacturing parameter constraints, such as avoiding hole overlap or too small hole diameters, we have introduced a penalty function similar to the one used in \Onlinecite{SehmiPRB17},
\be
P=\prod_i P_i\,, \mbox{with}\; P_i=\xi(\rh_i)\prod_{j \neq i} \xi(d_{ij})\,,
\ee
where $i$ is an index for the holes, 
\bea
\xi(x)     & = & \left \{
\begin{matrix}
	1           & x > x_0 \\
	1 + ((x-x_0)/x_{\mathrm{t}})^2  & x < x_0
\end{matrix} \right.\,,
\label{eqn:xi}
\eea
and $d_{ij}=|\xyh_i-\xyh_j|-\rh_i-\rh_j$ is the shortest distance between the rims of holes $i$ and $j$, and $\rh_i$ is the radius of hole $i$. We have used $x_0=20$\,nm, defining the lower limit of hole radius and wall thickness without penalty, and a response length $x_{\mathrm{t}}=5$\,nm at which the cost doubles when venturing beyond the limits. The quadratic cost increase (see \Eq{eqn:xi}) ensures that $F$ has a continuous derivative versus the hole size and positions, as required for the gradient descent minimisation.

\begin{figure}
	\includegraphics[width=\columnwidth]{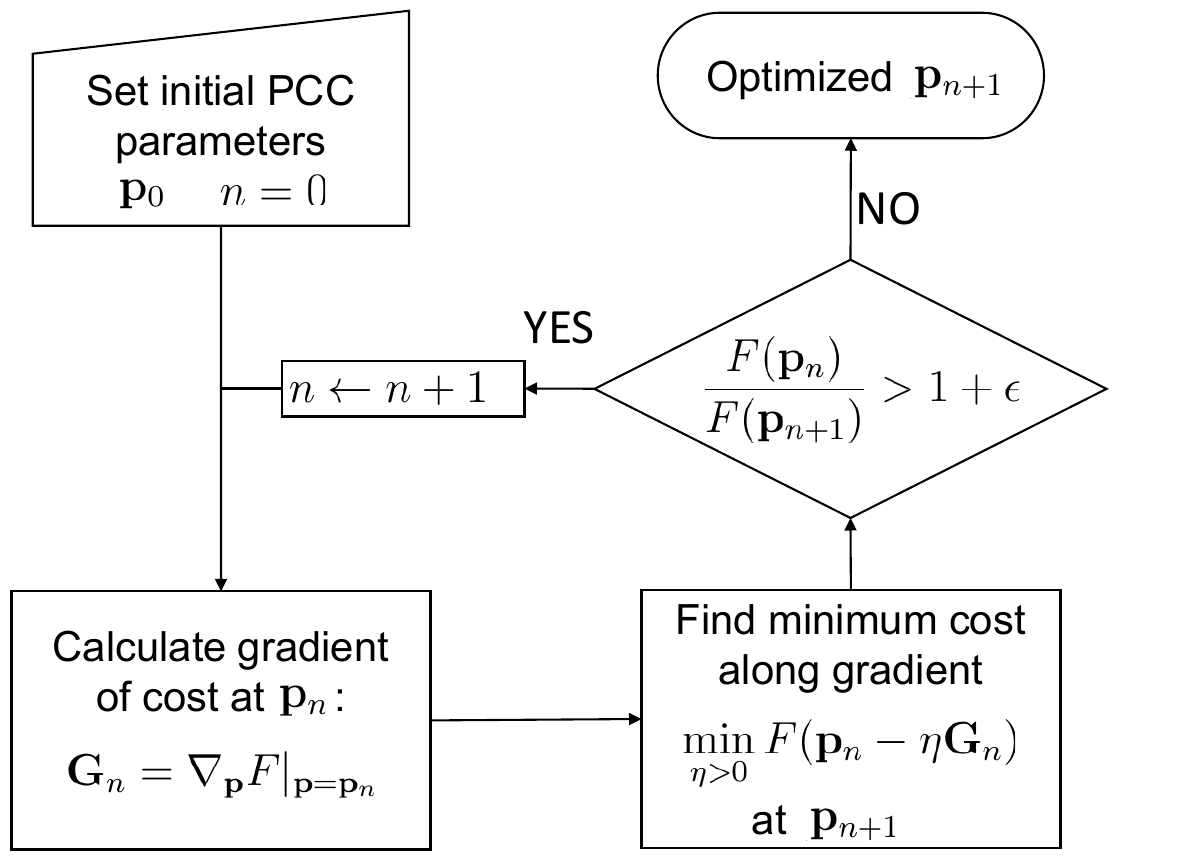}
	\caption{Flowchart of the iterative optimisation algorithm.}
	\label{fig:algorithm}
\end{figure}

\Fig{fig:algorithm} shows the flow diagram for the iterative optimisation of the cost function.
Starting from initial parameters $\parm_0$,  for example the nominal cavity, we calculate the gradient of the cost function $F(\parm)$ by simulating the cavity mode for small changes \parmd\ of the parameter values \parm, \xyhde\ for positions and \rhde\ for radii, chosen here to be $\xyhde=0.002\,a$ and $\rhde=0.001\,a$. These changes are small enough to neglect the influence of local curvature of the cost function, while large enough to limit the effect of simulation precision on the gradient, as detailed in \Sec{S-Gradient}. The gradient is calculated changing each parameter individually to assemble the gradient vector \Fgr. Notably, changing the geometry in COMSOL can lead to a change of the mesh structure, which affects the systematic simulation error, and can lead to a large error of the calculated gradient. To capture such cases, we calculate $F$ for parameter values of $\parm\pm\parmd$. If the obtained gradients for positive and negative changes are equal within 10\%, the average value is used. Otherwise, two additional calculations are performed at $\parm \pm 2\parmd$, providing a total of 4 gradient values from parameters separated by \parmd. As changes of the mesh structure occurs at discrete points during the geometry change, we can assume that it occurs at most once within the small geometric changes made, the value with the largest difference to the mean of the remaining values is discarded and the mean of the remaining three is used. We find that for the simulations done, such steps typically occur for 10\% to 50\% of the calculated gradients for changing individual parameters.
 
Having determined $\Fgr_n=\Fgr(\parm_n)$, the first minimum of $F(\parm)$ along $\Fgrh_n=\Fgr_n/|\Fgr_n|$ is determined as follows: $F$ is calculated for $\parm=\parm_n-m\zeta\Fgrh_n$, with $m=1,2,..$, and a step size $\zeta$ adaptively chosen to be small enough to provide an accurate quadratic interpolation, while large enough to reach the minimum in a small number of steps. We start the iteration with $2\zeta=F(\parm_n)/|\Fgr_n|$. To minimise computational time, we limit the search region in the COMSOL eigenmode solver to $\pm2$\% of the expected eigenfrequency and solve for a single eigenmode only. If one or more modes are returned by COMSOL within the search range, we compare the calculated field profiles with the last know one, e.g. obtained at the previous iteration. In order to follow the same mode during our iteration, we select the one with the highest correlation with the previous field, as detailed in \Sec{S-Corr}. If no modes with a correlation larger than 50\% are found within the range, we increase the search range by $\pm 2$\% and double the number of modes to be returned by the solver, until such a mode is found.  The expected eigenfrequency was taken as the one at $\parm_n$ when determining the gradient.
If $F$ decreases for $m=1$, we step along the gradient direction increasing $m$ until we pass the minimum in $F$, seen as an increase of $F$ from the previous step. Otherwise, we repeat halving $\zeta$ and recalculating $F$ until it decreases or $\zeta<0.01$. We then find the minimum position $\parm_{n+1}$ by a quadratic fit to $F$ as function of $m\zeta$ using the three lowest values of $F$ found across all points calculated along the gradient, including $\parm_n$. 
 
\begin{figure}
	\includegraphics[width=\columnwidth]{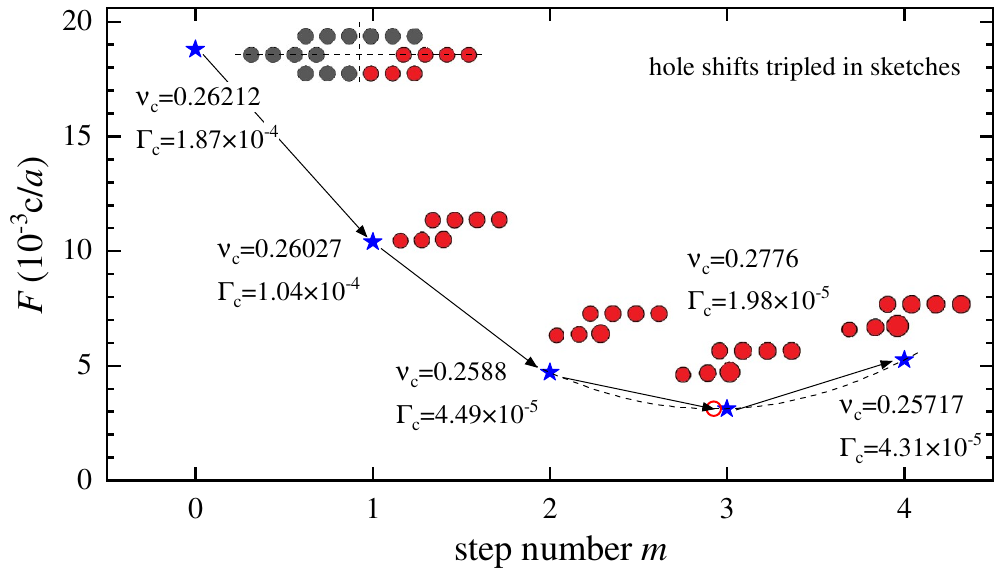}
	\caption{First minimisation iteration from a nominal L3 cavity, showing $F$ as function of step number $m$. The holes modified during the optimisation are sketched. For $m>0$ the holes defined by the two mirror symmetries of the structure indicated by dashed lines are omitted for clarity. The shifts of the holes have been tripled for clarity. The red circle indicates the position of the minimum cost function determined by a quadratic fit (dashed line) to the lowest three values of the cost function obtained during the minimisation.}
	\label{fig:descent}
\end{figure}

The gradient minimisation is then repeated starting from $\parm_{n+1}$, unless the relative change of $F$ from $\parm_n$ to $\parm_{n+1}$ is less than $\epsilon$, defining sufficient convergence to a local minimum, so that $\parm_{n+1}$ is accepted as the optimised position. We used $\epsilon=1\%$.
As example of a single gradient step we show in \Fig{fig:descent} the first iteration for an L3 cavity starting from the nominal geometry, using \nut=0.26 and $\alpha=100$. A five-fold reduction of the cost $F$ is achieved, and the illustrated change in the hole positions and sizes along the gradient shows that the corner hole is most relevant for this gradient, increasing in size and shifting towards the cavity. 

The gradient descent trajectory in the multidimensional \parm\ space depends on the starting point and the choice of $\alpha$. A low $\alpha=1$ results in a fast convergence to the target frequency \nut. Once approximately reached, $\alpha$ can be increased to give more weight to reducing the linewidth, but for large values the frequency deviation eventually will increase.  An algorithmic adjustment of $\alpha$ along the optimisation will be presented in a future work.

\section{Results}
\subsection{GaAs L3 cavity in water}
A nominal L3 PCC in a GaAs PC slab in water with lattice constant $a=325\nm$, hole radius $\rh=0.35a$, slab index $\nsl=3.4093$ and water index $\nme=1.3233+4.1918\times10^{-5}\mathrm{i}$ has a resonant mode at $\nuc=0.26212$ corresponding to $\lamc=1239.89$\,nm with $Q=1402$. We refer to this cavity without any optimisation of the position and size of the holes as nominal. 

\begin{figure}
	\includegraphics[width=\columnwidth]{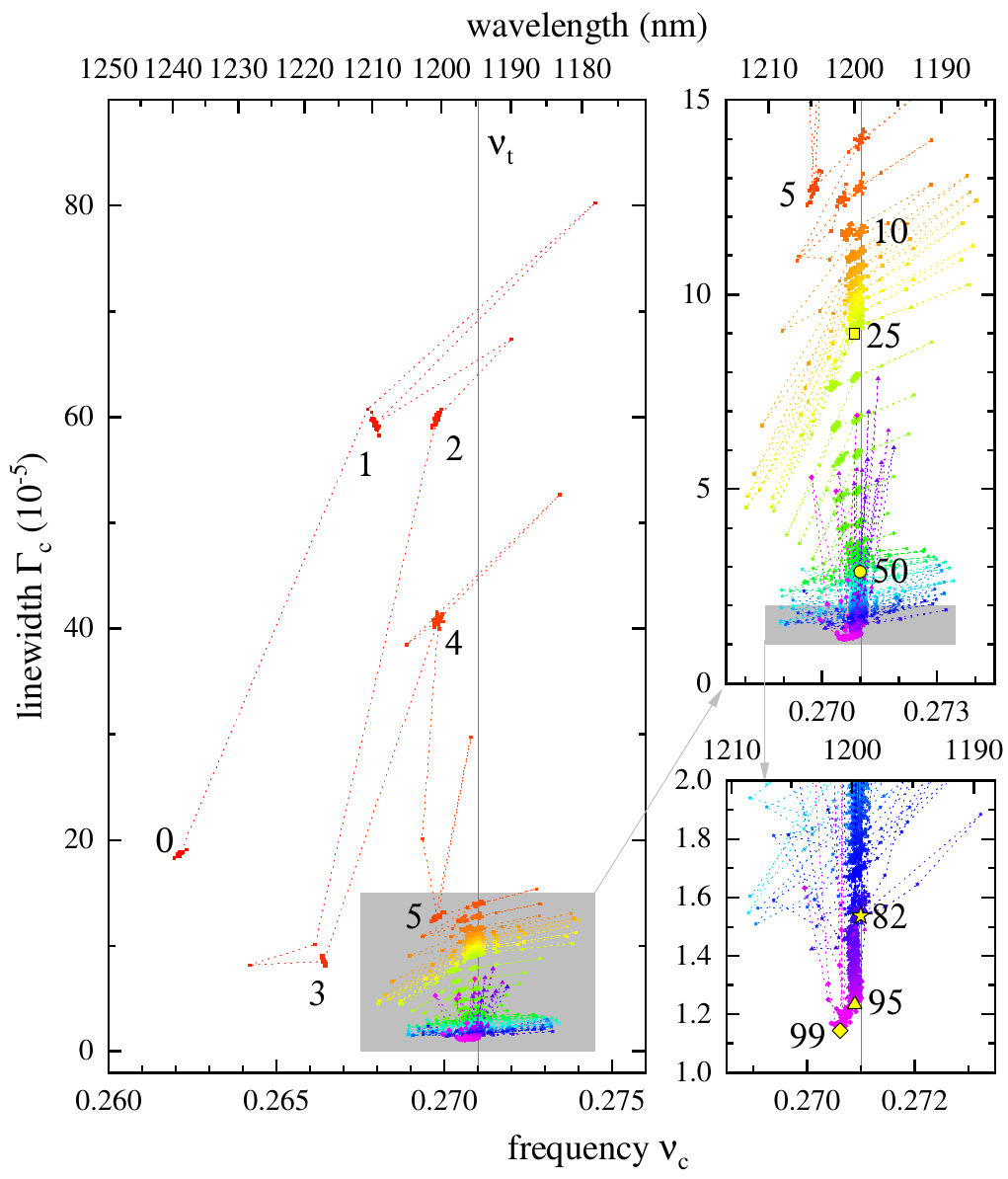}
	\caption{Optimisation trajectory for an L3 cavity starting at $\nuc=0.26212$ with unchanged remaining holes of the PC, to a target frequency of $\nut=0.27102$. The reduced linewidth $\gamcr$ is shown versus the reduced frequency \nuc\ for each simulation made. The clustered points represent the gradient calculations, which are numbered by the gradient step $n$ up to $n=5$ for clarity. The weighting $\alpha$ changes from 10 (squares), to 20 (circles), 40 (stars), 80 (triangles) and 160 (diamonds) for the final stage. The symbol colour hue changes for increasing simulation from red to green to blue to purple for clarity. The yellow larger symbols identify the end of the optimisation for the corresponding value of $\alpha$.}
	\label{fig:L3Opt_250THz}
\end{figure}

\Fig{fig:L3Opt_250THz} shows the evolution of the linewidth \gamcr\ and frequency \nuc\ of the cavity mode along the gradient descent simulations for $\nut=0.27102$, with each simulation done shown as symbol. Calculations of the gradient manifest as clusters of sympols, while the following steps along the descent are typically well separated. A weight $\alpha=10$ (square symbols) was used for the first tranche of iterations $n=0..25$. Initially, the detuning from \nut\ is the dominant contribution to the cost and the first two gradient steps $n=0,1$ mostly reduce the detuning. The next steps up to $n=5$ show a mix of changes, indicating the complex trajectory in the 16-dimensional parameter space \parm. For $n=6...24$ (see top right panel in \Fig{fig:L3Opt_250THz}), the even steps improve detuning while slightly deteriorating the linewidth, while the odd steps improve the linewidth while slightly deteriorating the detuning, with a pair reducing the linewidth by typically 3--6\%. This behaviour continues also after $\alpha$ is increased to 20 for $n=26..50$ (circles), with linewidth reductions up to 16\% every two steps. $\alpha$ is increased to 40 for $n=51..82$ (stars), over which \gamcr\ is halved, with \nuc\ keeping within 1\% from the target frequency. A further doubling of $\alpha$ to 80 for $n=83..95$ (triangles) provides another 20\% reduction in \gamcr\ ending at $\nuc=0.27089$ and $\gamcr=1.234\times10^{-5}$ (yellow triangle), corresponding to a $Q=\nuc/\gamcr=2.2\times10^4$, a sixteen fold increase from the nominal mode. Further doubling of $\alpha$ to 160 for $n=96..99$ (diamonds) increases $Q$ by some 10\%, but at the cost of increasing the detuning by a redshift of \nuc. This behaviour is a consequence of the  increasing losses with due to the expansion of the radiative cone with increasing frequency, rendering a larger \bkip\ range of the mode field emissive.

\begin{figure}
	\includegraphics[width=\columnwidth]{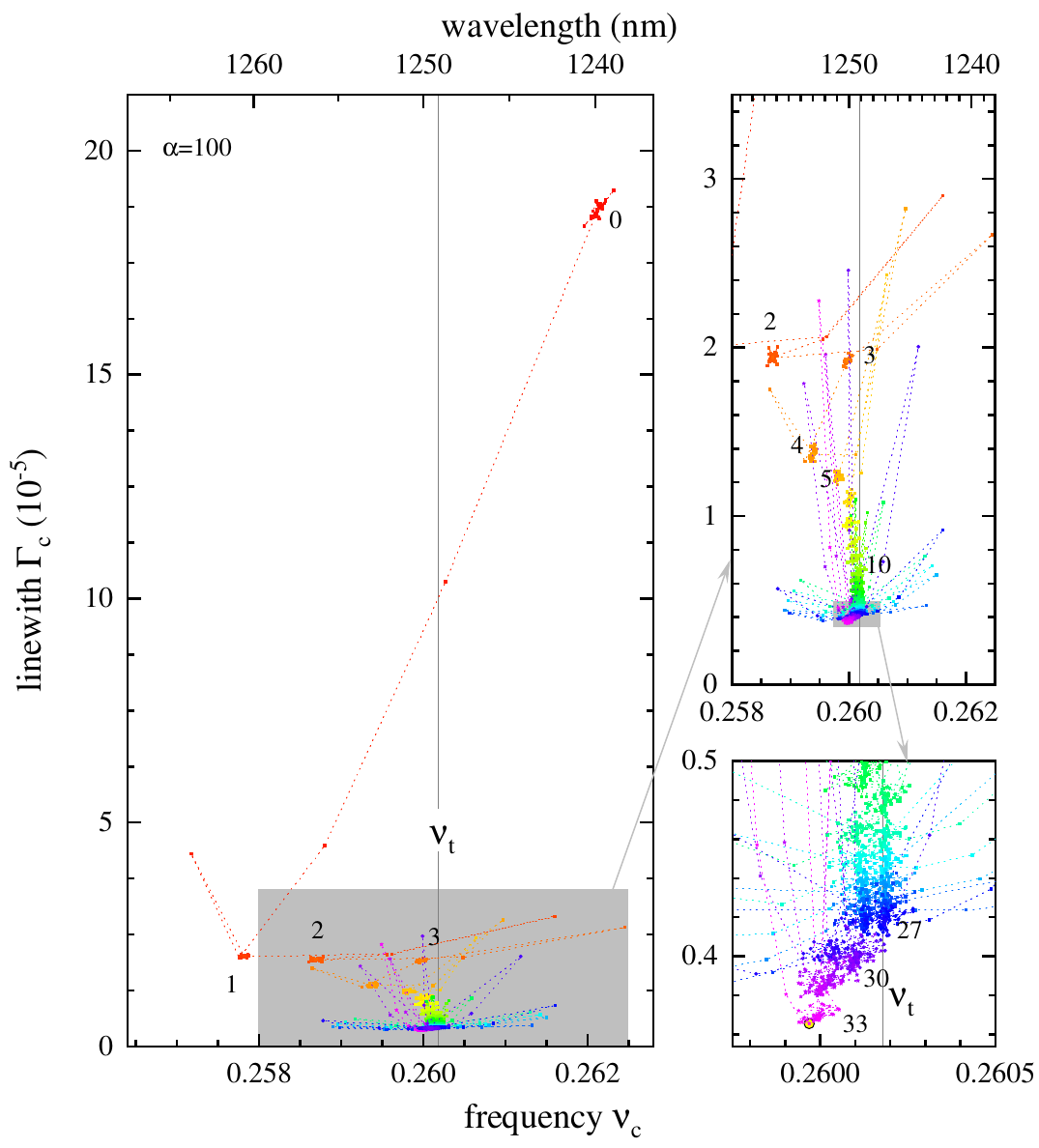}
	\caption{As \Fig{fig:L3Opt_250THz}, but for a target frequency $\nut=0.26018$. The weighting $\alpha$ changes from 100 (squares) for $n=0..26$ to 200 (circles) for $n=27..30$, to 400 (stars) for $n=31..33$.}
	\label{fig:L3Opt_240THz}
\end{figure}

Using instead a target frequency $\nut=0.26018$, close to the one of the nominal L3 cavity, the optimisation trajectory is given in \Fig{fig:L3Opt_240THz}, starting with $\alpha=100$.  Here, already at the first step in the descent, we observe a ten-fold reduction of linewidth. For $n=1..6$ the optimisation moves the frequency closer to \nut\ while gradually also reducing the linewidth. The following steps up to $n=26$ show, as in \Fig{fig:L3Opt_250THz}, an alternating reduction of linewidth and detuning, yielding overall a two-fold decrease of the linewidth. At $n=27$, we change the weighting $\alpha$ to 200, and yielding a further reduction of \gamcr, but with a clear detuning from \nut, as can be expected for large $\alpha$. For $n=31..33$ we further increased $\alpha$ to $400$, resulting in an increasing detuning and a further reduction of linewidth by about 10\%, and a final optimised eigenmode with $\nuc=0.25997$ and $\gamcr=3.65\times10^{-6}$, corresponding to a $Q=7.12\times10^4$, a fifty-fold increase from the nominal cavity. The detuning to \nut\ is $-2.1\times10^{-4}$, about 200 times larger than the linewidth, in keeping with the value of $\alpha$ used. Again, we find that decreasing the frequency allows decreasing the linewidth, as expected from the decrease of radiative cone size.

\begin{figure}
	\includegraphics[width=\columnwidth]{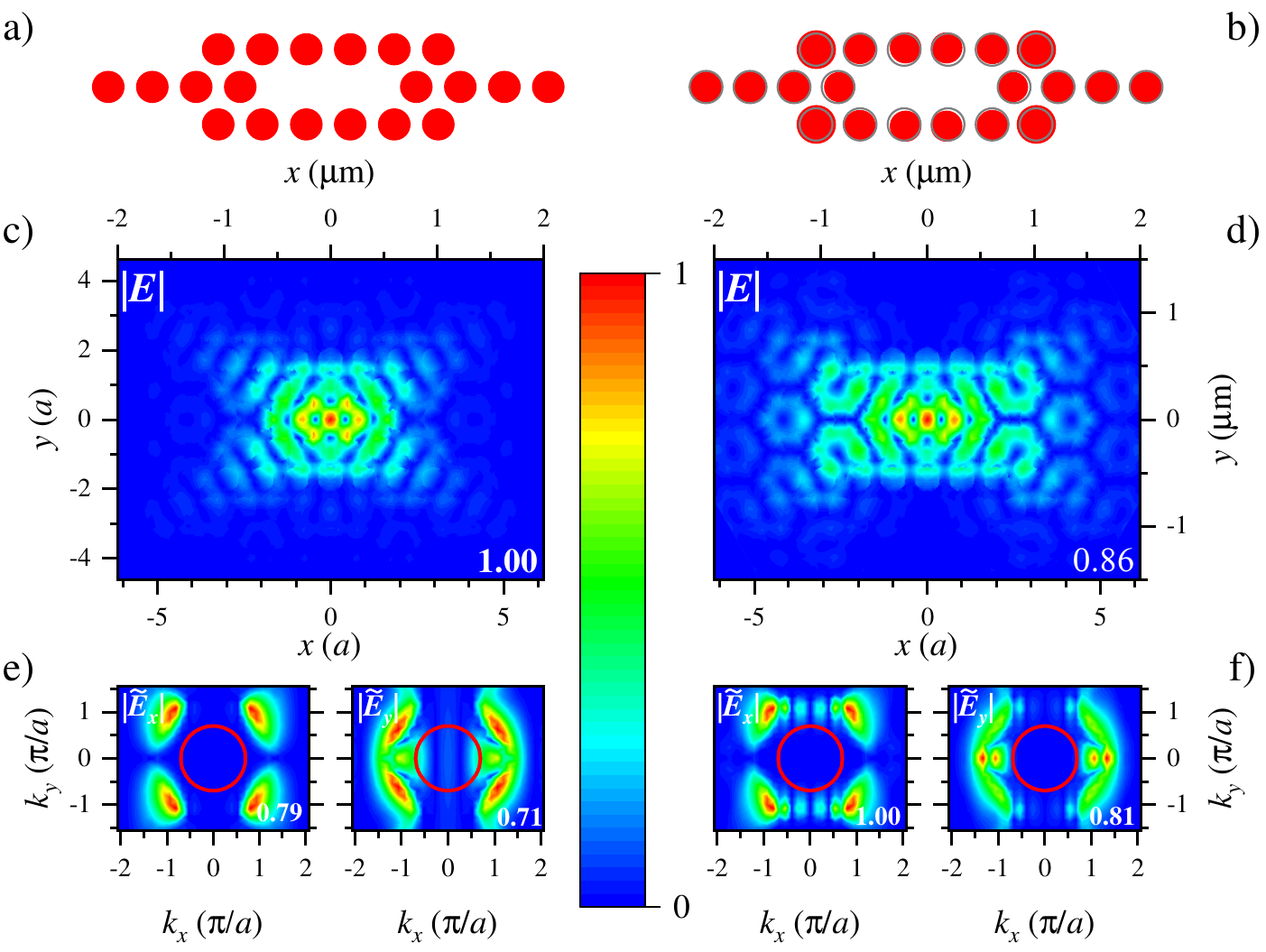}
	\caption{Comparison of the nominal L3 cavity (left) and the optimised cavity with $\nut=0.26018$ (right). The geometries of the parametrised holes surrounding the cavity are shown in a) and b) with grey circles indicating the nominal geometry. The magnitude of the electric field of the RS at the central plane of the slab is shown in c) and d) in real space. The corresponding $x$ and $y$ polarised field components in \bkip\ space are shown in e) and f), with the edge of the radiative cone $|\bkip|=\nme\omc/c$ shown as red circles. The colour scale is given, and data are shown normalised to their maximum. Relative values of maxima are given.}
	\label{fig:L3Mode}
\end{figure}

The hole structure for the nominal and the optimised L3 of \Fig{fig:L3Opt_240THz} is given in \Fig{fig:L3Mode}(a,b). The dominant change is similar to the one seen in the initial iteration \Fig{fig:descent}, the radius increase of the corner hole. Additionally, the inner holes are slightly pushed away from the cavity. Notably, the in-line holes are only slightly changed. The values of the optimised parameters are summarised in \Tab{S-tab:L3ParamOpt}, with their evolution during optimisation shown in \Fig{S-fig:L3ParamOpt}. The RS field for the nominal and optimised L3 are shown in \Fig{fig:L3Mode}(c,d) in real space and in \Fig{fig:L3Mode}(e,f) in \bkip\ space. In real space, the optimisation resulted in a slightly more extended mode and a more hexagonal rather than stripy field pattern in the PC outside the cavity. In \bkip\ space, the $x$ polarised component $E_x$ has already in the nominal L3 cavity negligible field in the light cone. $E_y$ instead has significant overlap close to the edges of the cone in $k_x$ direction, as well a weaker stripe along $k_y$ at $k_x$ around zero. In the optimised L3, the latter has vanished, while the former has been pushed out to larger $|k_x|$, just outside the light cone, thus suppressing radiative losses to enable the high $Q$. Clearly, this remaining overlap is susceptible to the light cone size, providing the mechanism by which a frequency reduction decreases the radiative losses seen in \Fig{fig:L3Opt_240THz}.  

\begin{figure}
	\includegraphics[width=\columnwidth]{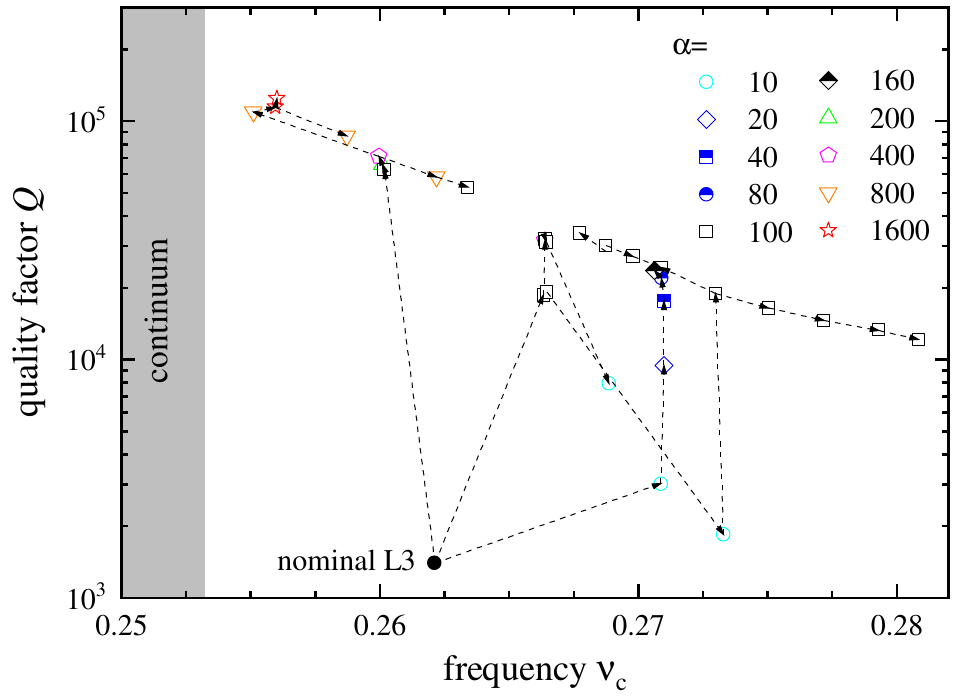}
	\caption{Quality factor $Q$ of optimised L3 cavities versus their frequency \nuc, for different values of \nut\ and different starting parameters. The shaded area is below the PC band gap. The filled circle refers to the nominal cavity, while the empty symbols refer to optimisation results for different values of \nut. The shape and colour of the symbols encodes the value of $\alpha$ used (see legend). The arrows point from the initial to the final mode of each optimisation.}
	\label{fig:L3Summary}
\end{figure}

We found that to reduce the compute time of the optimisation, and possibly find lower local minima of the cost function, an optimised cavity with high $Q$ and eigenfrequency near \nut\ can be used as starting point. For example, a $Q$ exceeding $10^5$ is reached for $\nut=0.25511$ starting with the optimised solution at $\nuc=0.25997$. \Fig{fig:L3Summary} shows a summary of the highest $Q$ achieved so far in our optimisations for different values of \nut. We find that for $\nut>0.282$, the optimisation returns final cavities with significant detuning and $Q<10^4$, indicating that the radiative cone is too large for the mode to avoid it given the constraints of the hole geometry used. The results of the different optimisations are tabulated in \Tab{S-tab:L3Summary}. However, we emphasize that it is not excluded that geometries with higher Q exist even with the geometry constraints, but are not found along the gradient descent. Starting from a large number of random parameters within the constraints could be investigated to improve on this. 

\subsection{GaAs H1 cavity in water}
A H1 cavity in the same PC as above has a six-fold symmetry and thus two degenerate resonant modes\,\cite{MinkovSR14} due to the two polarisation states, at $\nuc=0.28541$. The slight symmetry breaking introduced by the meshing results in RSs M1 (M2) with polarisations aligned with the horizontal (vertical) axis of the domain, respectively. 
In the optimisation, the six-fold rotation and reflection symmetry was kept and the first three hole rows surrounding the omitted hole were varied, so that \parm\ consists of the radial shifts and radius changes of each row. In the optimisation, we used M1, and due to the symmetry the M2 frequency is equal to M1, and we find a splitting around 0.1\% from symmetry breaking by meshing. Similarly, the $Q$ of M1 and M2 are equal within 2\%. The nominal H1 cavity has a M1 mode of low $Q=127$. Starting from the nominal H1 cavity, optimising for $\nut=0.27102$ resulted in $Q=5.9\times10^4$, a four hundred sixty fold increase.

\begin{figure}
	\includegraphics[width=\columnwidth]{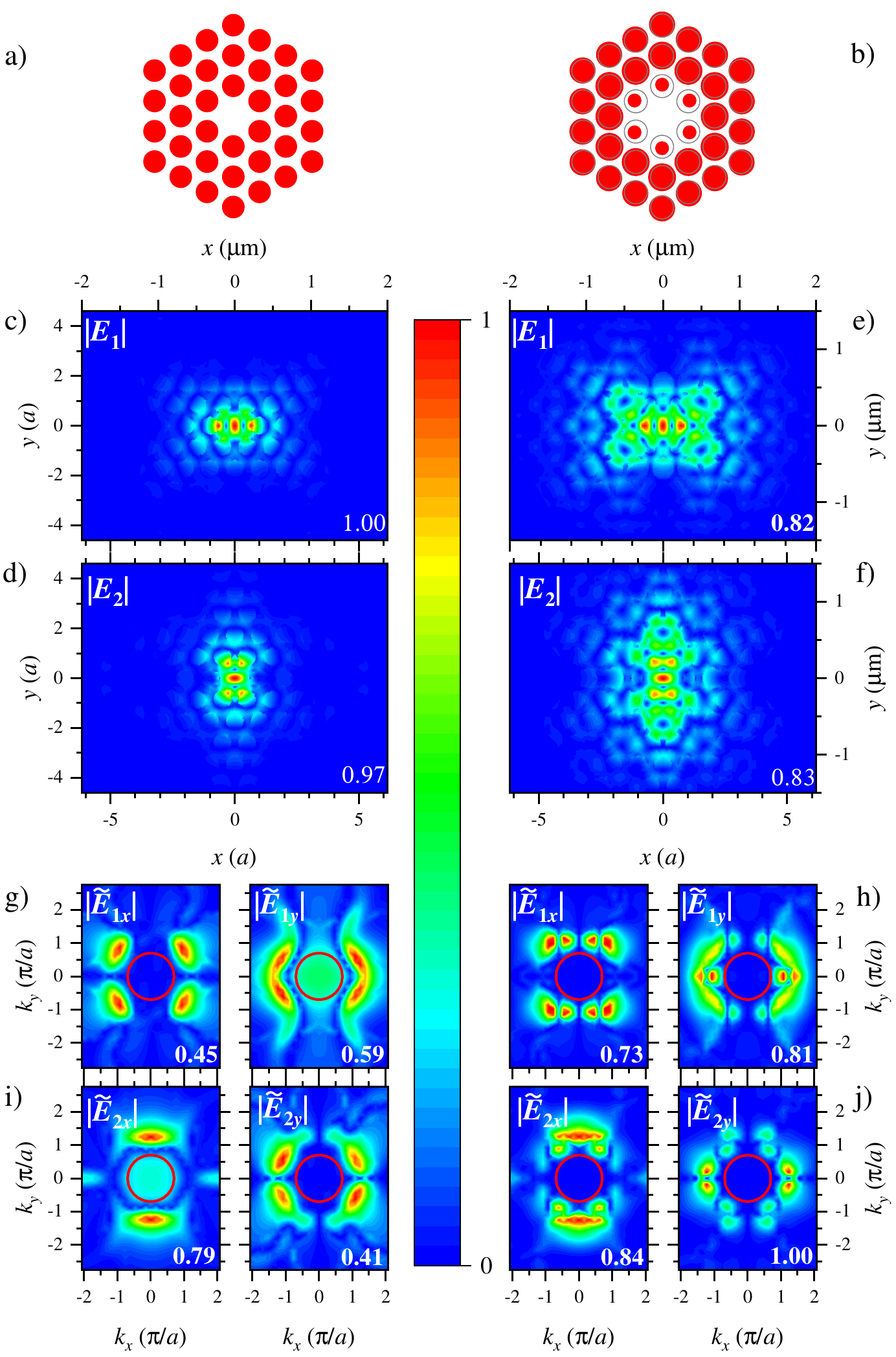}
	\caption{Comparison of the nominal H1 cavity modes (left) and the optimised cavity with $\nut=0.27083$ (right). The geometries of the holes surrounding the cavity are shown in a) and b) with grey circles, respectively. The magnitude of the electric field of the  M1 and M2 modes at the central plane of the slab is shown in real space in c) and d) for the nominal and in e) and f) for the optimised cavity. The corresponding $x$ and $y$ polarised field components in \bkip\ space are shown in g) to j), with the edge of the radiative cone shown as red circles. The colour scale is given, and data are shown normalised to their maximum. Relative values of maxima are given.
	}
	\label{fig:H1Mode}
\end{figure}

The corresponding hole structure is given in \Fig{fig:H1Mode}(a,b) for the nominal and optimised H1. The dominant change is the decrease of hole size in the first row and increase in the second row and a smaller increase in the third row. The corresponding optimised parameters are given in \Tab{S-tab:H1ParamOpt}. The RS field is shown in \Fig{fig:H1Mode}(c,d) in real space and in \Fig{fig:H1Mode}(e,f) in \bkip\ space. In real space, the optimised H1 shows a more extended field and again roughly hexagonal field patterns as in the L3 case. In \bkip\ space, the nominal H1 cavity shows strong field components in the light cone peaked at $\bkip=0$, in $E_y$ for M1 and in $E_x$ for M2, which is the reason for the poor quality factor.
This feature emerges from the dominant single anti-node of the field at the cavity center, which is not compensated by the out-of phase (see the mode phase shown in \Sec{S-sec:H1 Phase}) antinodes to the left and right of the centre. In the optimised H1, the latter have increased strength to achieve complete destructive interference at $\bkip=0$.

\begin{figure}
	\includegraphics[width=\columnwidth]{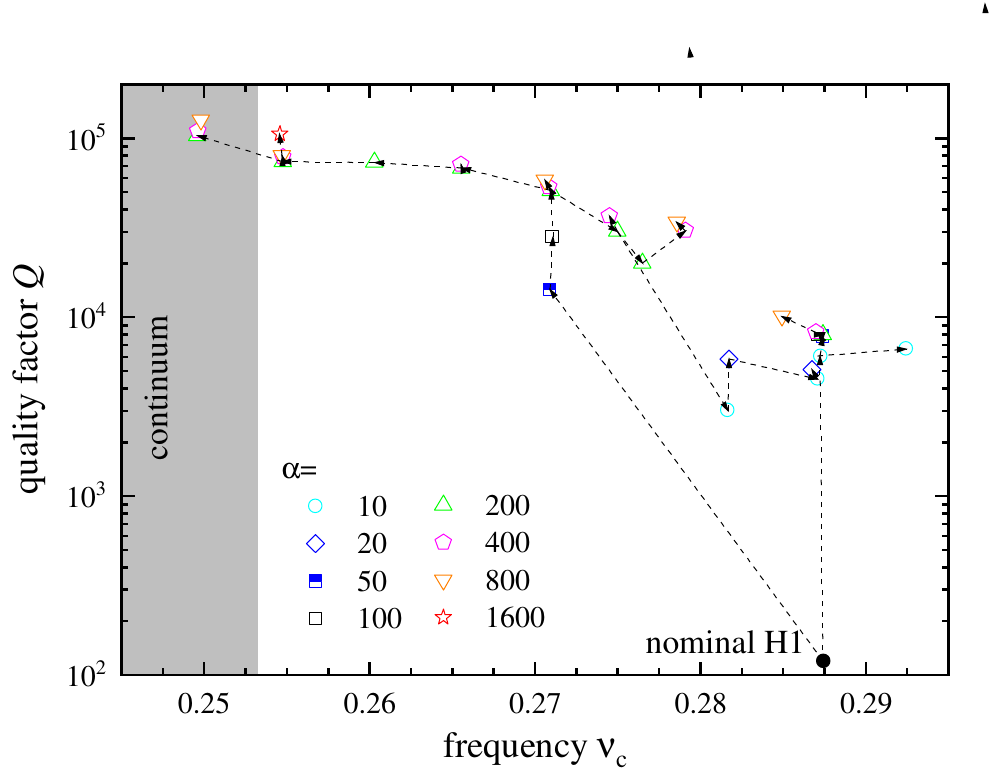}
	\caption{As \Fig{fig:L3Summary} for the H1 cavity. }
	\label{fig:H1Summary}
\end{figure}

An approximately 400-fold increase of $Q$ to above $5\times10^4$ was found for $\nuc=0.255 ... 0.280$ (see \Fig{fig:H1Summary}), corresponding to an about $50$\,nm wide range of cavity mode wavelengths. As observed in the L3 cavity case, the highest $Q$ are achieved for optimised cavities with frequencies towards to the lower edge of the photonic band gap. This is related to the smaller radiative cone as discussed before, but also to the longer decay length in the PC, reducing broadening of the localised cavity mode in \bkip\ space. However, the longer decay length also increases the mode volume, which is reducing the coupling strength in cavity QED for Purcell enhancement and strong coupling. It also requires larger PC slabs, which is detrimental for high density integration of cavities. We thus evaluated the mode volume, using the simplified expression valid for high $Q$ modes\cite{MuljarovPRB16a} 
\be 
V_{\rm m}=\frac{\int{\epsilon(\br)|E(\br)|^2d\br}}{\epsilon(\br_0)|E(\br_0)|^2}\,,
\ee
with the center of the cavity $\br_0$, and integration over the slab and medium domains.  We found an about two-fold increase from the nominal H1 cavity ($\Vm=0.0066\,$\textmu m$^3$) to the optimised cavity of \Fig{fig:H1Mode} ($\Vm=0.0167\,$\textmu m$^3$). For the L3 cavity the optimisation increases the mode volume by only about 30\% ($\Vm=0.017\,$\textmu m$^3$ for the nominal, compared to $\Vm=0.022\,$\textmu m$^3$ for the optimised cavity of \Fig{fig:L3Mode}). The increase in $Q$ factor is thus much larger than the increase of mode volume, so that the optimisation is highly beneficial for Purcell enhancement and strong coupling of quantum emitters.

%\Fig{fig:H1Opt} shows the intermediate steps of the different optimisations, with the optimised solutions indicated by black symbols. The figure shows that faster convergence can be obtained if optimised solutions at nearby frequencies are used as start point.

 %\begin{figure}
 %	\includegraphics[width=\columnwidth]{H1Opt}
% 	\caption{Optimisation trajectories for different target frequencies for a H1 cavity of GaAs in water. The colour encodes the optimisation step (from red to purple). {\bf FM: same comment as above: the optimisation at 0.282 is with a different refractive index with respect to the others. I have included another optimisation at 0.28728, with the same refractive index, in case we want to show all results with the same index.} {\bf WL: add target frequencies as arrows or dashed lines}}
% 	\label{fig:H1Opt}
% \end{figure}

\subsection{Si L3 cavity in air}
\begin{figure}
	\includegraphics[width=\columnwidth]{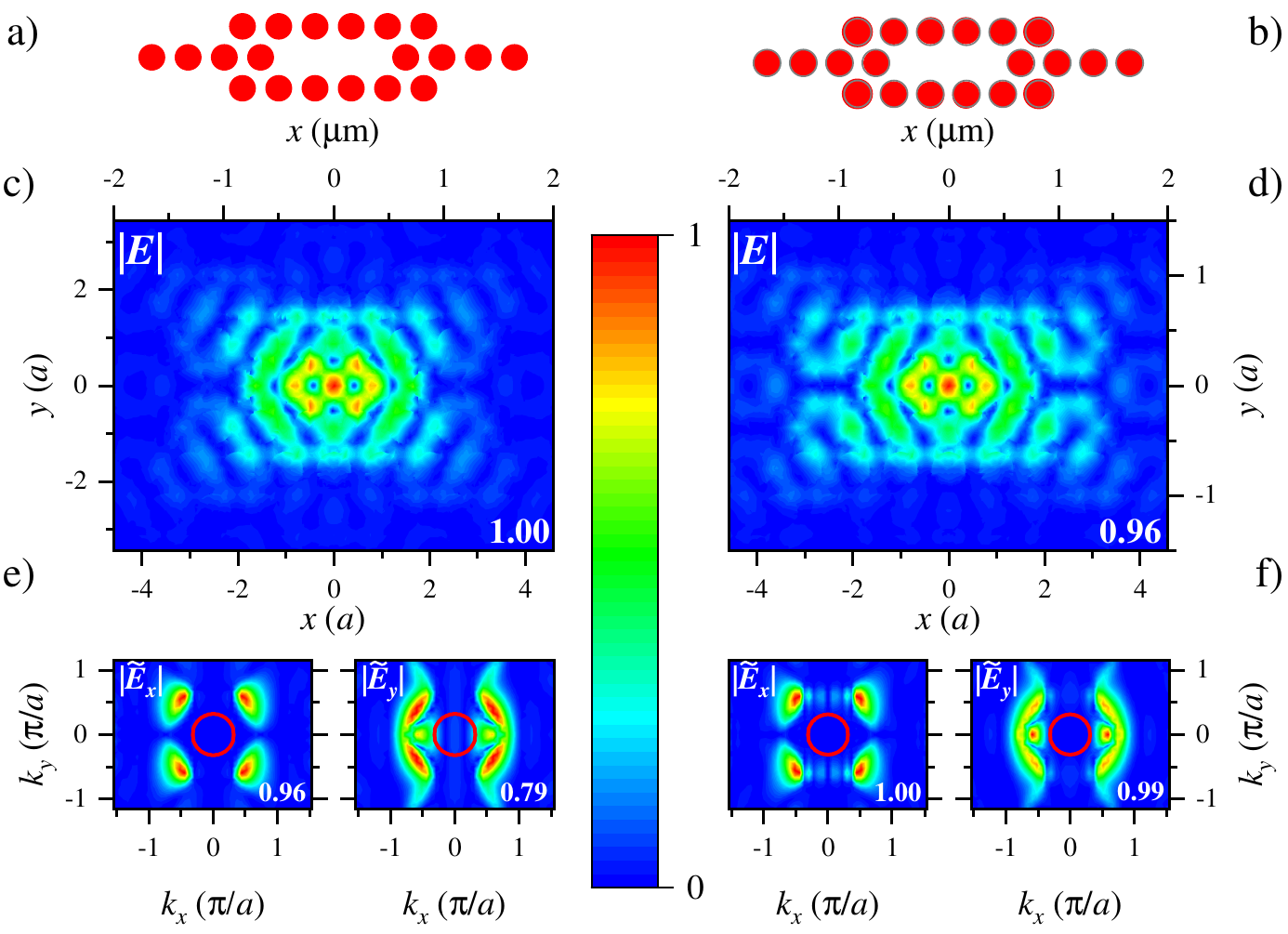}
	\caption{As \Fig{fig:L3Mode} but for a silicon L3 cavity in air with $\nut=0.2828$.}
	\label{fig:SiL3Mode}
\end{figure}

To compare with published high-Q geometries, we show here the optimisation of a L3 cavity using a Si slab in air. Using a lattice costant $a=437$\,nm, a slab thickness of $0.55a=240$\,nm and a hole radius of $0.35a=153$\,nm, the nominal L3 has a resonant mode at $\nuc=0.28212$ corresponding to a resonant wavelength $\lamc$ around 1.55\,\textmu m, in the fiber-optic C-band. The mode shows a $Q$ of 4181. The optimisation with $\nut=0.2828$\,\ca\  and $\alpha=100..200$ produces an optimised mode with $Q$ exceeding $2.5\times10^5$ (see \Fig{fig:SiL3Mode}), limited by the meshing used in the simulation ( finer meshing of the hole edges - from 8 to 12 segments - yielded an optimised Q-factor of $3.2\times10^5$.) Similarly to what observed for the case of the optimisation of a L3 cavity slab made of GaAs in water, the biggest change observed is on the size of the corner holes. 
Also the mode field in $\bkip$ space shows a similar behaviour with the peak in the $E_x$ component close to the edge of the radiative cone being pushed to higher $k_y$, reducing its overlap with the radiative cone. The obtained parameters are summarised in \Sec{S-sec:SiL3ParamOpt}.
In this case the optimisation approximately doubles the mode volume ($\Vm=0.017\,$\textmu m$^3$ for the nominal, compared to $\Vm=0.032\,$\textmu m$^3$ for the optimised cavity of \Fig{fig:SiL3Mode}).

\section{Conclusions}
We have demonstrated an algorithm to design a photonic crystal cavity supporting a resonant mode at a given target frequency with minimum linewidth, and thus maximum quality factor. The method is based on a gradient descent approach where the position and size of the holes surrounding the cavity are changed in an iterative way, considering physical constraints and required symmetries. This allows to design cavities in a photonic crystal slab with resonances at different frequencies and consistently high $Q$. We have also elucidated the physical loss mechanism as emission into the radiative cone, and how the optimisation is redistributing the mode field in wavevector space to avoid this cone. Importantly, it is of general use for arbitrary cavity geometries and materials, and we provide the source code to enable widespread application across different field including sensing, telecom, and quantum technology.

\section*{Acknowledgements}
N.M. acknowledges support for her PhD studies by the EPSRC DTP [grant n. 2279386]. F.M. acknowledges the Ser Cymru II programme (Case ID 80762-CU-148) which is part-funded by Cardiff University and the European Regional Development Fund through the Welsh Government. The authors would like to thank Dr Daryl Beggs for helpful discussions.

%\nocite{*}
\section*{References}
%\bibliography{langsrv,PCCOpt}% Produces the bibliography via BibTeX.

%aipnum4-2.bst 2019-01-14 (MD) hand-edited version of apsrev4-1.bst

\end{document}